\documentstyle[12pt]{article}
\begin{document} 
\newcommand{\first}{1^{\hbox{\em \`ere}}}
\newcommand{\secnd}{2^{\hbox{\em \`eme}}}
\newcommand{\vect}[1]{{\bf #1} }
\newcommand{\bra}[2]{\langle #1,#2 \vert}
\newcommand{\ket}[2]{\vert #1,#2 \rangle}
\newcommand{\vet}[2]{\vert #1,#2)}
\newcommand{\vra}[2]{(#1,#2 \vert}
\newcommand{\cg}[8]{(#1,#2,#3,#4 \vert #5,#6,#7,#8)}
 \newcommand{\qpn}[1]{[[#1]]_{qp}}
 \newcommand{\ti}[1]{\tilde #1} 
\newcommand{\somn}{\sum_{n=0}^{\infty}}
\newcommand{\somk}{\sum_{k=0}^{\infty}}
\newcommand{\somp}{\sum_{p=0}^{n}}
\newcommand{\somm}{\sum_{m=0}^{\infty}}

% BEGINNING OF THE FILE

\vskip 0.25 true cm

\centerline {\bf A SYSTEM OF INTEREST IN SPECTROSCOPY:}

\centerline {\bf THE $qp$-ROTOR SYSTEM}
\vskip 1 cm
\centerline {R\'emi BARBIER and Maurice KIBLER}
\vskip 0.5 cm
\centerline {Institut de Physique Nucl\'eaire de Lyon, 
IN2P3-CNRS et Universit\'e Claude Bernard,} 

\centerline {43 Boulevard du 11 Novembre 1918, 
             F-69622 Villeurbanne Cedex, France}
\vskip 1 cm

\centerline {\bf Abstract}

\vskip 0.3 true cm

\noindent 
A rotor system, having the symmetry afforded by the 
two-parameter quantum algebra $U_{qp}({\rm u}_2)$, is
investigated in this communication. This system is useful 
in rotational spectroscopy of molecules and nuclei. In 
particular, it is shown to lead to a model 
(viz., the $qp$-rotor model) for describing 
(via an energy formula and a $qp$-deformation 
of E2 reduced transition probabilities) rotational 
bands of deformed and superdeformed nuclei. 

\vskip 1 true cm 

\baselineskip = 0.55 true cm
 
\section{Introduction}
Quantum groups and quantum algebras, introduced at the 
beginning of the eightees,$^{1-5}$ 
continue to attract much attention both in mathematics and 
physics. 
For the Physicist, a quantum algebra is commonly considered as a deformation 
($q$-deformation) of a given Lie algebra.
During the last four years, several works 
 have been performed on two-parameter quantum algebras and 
quantum groups ($qp$-deformations).$^{6-17}$ 

Most of the physical applications, ranging from chemical 
physics to particle physics, have been mainly concerned up to 
now with one-parameter quantum algebras ($q$-deformations). In 
particular, in nuclear physics we may mention applications to 
rotational spectroscopy of deformed and superdeformed nuclei,$^{18-26}$ 
to the interacting boson model,$^{27,28}$ 
to the Moszkowski        model,$^{29,30}$ 
to the U(3) shell        model$ ^{31}$ 
and to the 
Lipkin-Meshkov-Glick     model.$^{32}$ There exist also applications 
to particle physics, as for example to quote a few, to hadron 
mass formulas$^{33,34}$ 
and to Veneziano amplitudes.$^{35,36}$ Among 
the just mentioned applications, only the ones in 
Refs.~25 and 36 rely on the use of 
two-parameter deformations. 

The aim of the present communication is 
to describe a rotor system, 
with the $U_{qp}({\rm u}_2)$ symmetry, 
that constitutes a basic tool for the nonrigid rotor 
model briefly introduced in Ref.~25. The latter model, 
referred to as the $qp$-rotor model, is based on the 
two-parameter quantum algebra $U_{qp}({\rm u}_2)$ 
in contradistinction with the 
$q$-rotor models introduced by Iwao$^{18}$ and 
Raychev, Roussev and Smirnov$^{19}$ 
(see also Refs.~20-23 and 26) 
that are based 
on the one-parameter quantum algebra $U_{q}({\rm su}_2)$. 

The organization of this work is as follows. The 
 $qp$-rotor model is introduced in Sec.~2. 
Subsection 2.1 deals with the 
mathematical ingredients of the model. The $qp$-rotor model 
itself is developed in Subsec.~2.2 (rotational energy formula) 
and in Subsec.~2.3 (E2 reduced transition probabilities). 
Finally, some concluding remarks are presented in Sec.~3. 

\section{A $qp$-Rotor System}
 
\subsection{The quantum algebra $U_{{qp}}({\rm u}_2)$}
The             quantum algebra $U_{{qp}}({\rm u}_2)$ can be 
constructed from two pairs, say $\{ \ti{a}_+^+ ,\ti{a}_+ \}$ and 
                                $\{ \ti{a}_-^+ ,\ti{a}_- \}$, of 
$qp$-deformed (creation and annihilation) boson 
operators. The action of these $qp$-bosons 
on a nondeformed two-particle Fock space 
$ \{ \ket{n_+}{n_-} : n_+ \in {\bf N} , \ 
                      n_- \in {\bf N}  \} $ 
is controlled by   
 \begin{eqnarray}
\ti{a}^+_+\;  \ket{n_+}   {n_-} & = & 
              \sqrt{ \qpn{n_+ + \frac{1}{2} + \frac{1}{2}}} \; 
              \ket{n_+ +1}{n_-} ,   \nonumber \\
\ti{a}_+  \;  \ket{n_+}   {n_-} & = & 
              \sqrt{ \qpn{n_+ + \frac{1}{2} - \frac{1}{2}}} \;  
              \ket{n_+ -1}{n_-} ,   \nonumber \\
\ti{a}^+_-\;  \ket{n_+}   {n_-} & = & 
              \sqrt{ \qpn{n_- + \frac{1}{2} + \frac{1}{2}}} \; 
              \ket{n_+}{n_- +1} ,   \nonumber \\
\ti{a}_-  \;  \ket{n_+}   {n_-} & = & 
              \sqrt{ \qpn{n_- + \frac{1}{2} - \frac{1}{2}}} \;  
              \ket{n_+}{n_- -1} .   
\label{eq:qpbo}   
\end{eqnarray}
% (1)
In the present work, we use the notations
\begin{equation}
\qpn{X} : = { \frac{q^X - p^X}{q-p}}
\label{eq:not1}
\end{equation}
% (2)
and
\begin{equation}
[X]_q : = [[X]]_{qq^{-1}} = {\frac{q^X - q^{-X}}{q-q^{-1}}} ,
\label{eq:not2}
\end{equation}
% (3)
where $X$ may stand for an operator or a (real) 
number. For Hermitean conjugation requirements, the values of 
the parameters  
$q$ and $p$ must be restricted to some domains that can be classified 
as follows: (i) $q \in {\bf R}$ and 
                 $p \in {\bf R}$, (ii) 
                 $q \in {\bf C}$ and 
                 $p \in {\bf C}$ with $p=q^{\ast}$ 
(the ${\ast}$ indicates complex conjugation), and (iii)
                 $q = p^{-1} = {\rm e}^{{\rm i} \beta}$ with 
$0 \le \beta < 2 \pi$. The two pairs  $\{ \ti{a}_+^+ ,\ti{a}_+ \}$ and  
                                        $\{ \ti{a}_-^+ ,\ti{a}_- \}$
of $qp$-bosons commute and satisfy 
\begin{equation}
 \ti{a}  _{\pm} \ti{a}^+_{\pm}  =  \qpn{N_{\pm}+1} , \qquad 
 \ti{a}^+_{\pm} \ti{a}  _{\pm}  =  \qpn{N_{\pm}  } , 
\label{eq:commu}
\end{equation}
% (4)
where $N_{+}$ and $N_{-}$ are the usual number operators with 
\begin{equation}
N_{\pm} \ \ket{n_+}{n_-} = 
n_{\pm} \ \ket{n_+}{n_-} .
\label{eq:openomb}  
\end{equation}
% (5)
Of course, the $qp$-bosons $\ti{a}^+_{\pm}$ and 
                           $\ti{a}  _{\pm}$ 
reduce to ordinary bosons (denoted as 
                           $   {a}^+_{\pm}$ and 
                           $   {a}  _{\pm}$
in Refs.~37 and 38 and 
in Subsec.~2.3) in the limiting situation where $p = q^{-1} \to 1$. 

The passage from the (harmonic oscillator) state vectors 
$\ket{n_+}{n_-}$          to angular momentum state vectors $\vet{I}{M}$
is achieved through the relations
\begin{equation}
 I := \frac{1}{2}(n_+ + n_-), \quad \qquad 
 M := \frac{1}{2}(n_+ - n_-)  
\label{eq:sch1}
\end{equation}
% (6)
and
\begin{equation}
 \vet{I}{M}\  \equiv \ \ket{I+M}{I-M} \  = \  \ket{n_+}{n_-}. 
\label{eq:sch2}
\end{equation} 
% (7)
Equations (\ref{eq:qpbo}) may thus be rewritten as 
 \begin{eqnarray}
\ti{a}_{\pm}^+ \; \;  \vet{I}{M} & = & 
\sqrt{ \qpn{I{\pm}M+\frac{1}{2}+\frac{1}{2}}}\; \;  
\vet{I+\frac{1}{2}}{M{\pm} \frac{1}{2}} ,   \nonumber \\
\ti{a}_{\pm}   \; \;  \vet{I}{M} & = & 
\sqrt{ \qpn{I{\pm}M+\frac{1}{2}-\frac{1}{2}}}\; \;
\vet{I-\frac{1}{2}}{M{\mp} \frac{1}{2}} ,     
\label{eq:qpbo2}  
\end{eqnarray}
% (8)
so that the $qp$-bosons behave as ladder operators for the 
quantum numbers $I$ and $M$ (with $\vert M \vert \le I$). 
 
We are now in a position to introduce  a $qp$-deformation of the 
Lie algebra ${\rm u}_2$. A simple calculation 
shows that the four operators $J_\alpha$ ($\alpha = 0,3,+,-$) 
given by
\begin{equation}
J_0 \  := \ {{1}\over{2}}(N_+ + N_-) , \quad
J_3 \  := \ {{1}\over{2}}(N_+ - N_-) , \quad 
J_+ \  := \  \ti{a}_+^+ \ti{a}_-     , \quad 
J_- \  := \  \ti{a}_-^+ \ti{a}_+     
\label{eq:gener1} 
\end{equation}
% (9)
satisfy the following commutation relations$^{14,39}$ 
 \begin{equation}
[J_3,J_\pm   ] \ = \ \pm J_\pm ,  \qquad
[J_+,J_-     ] \ = \ (qp)^{J_0-J_3} \ \qpn{2J_3} ,  \qquad
[J_0,J_\alpha] \ = \ 0.
\label{eq: comm}
 \end{equation}
% (10)
We refer to $U_{ {qp}}({\rm u}_2)$ the (quantum) algebra 
described by (\ref{eq: comm}). To endow $U_{ {qp}}({\rm u}_2)$ with a Hopf 
algebraic 
structure, it is necessary to introduce a co-product $\Delta_{qp}$. 
The latter co-product is such that:$^{14}$ 
\begin{eqnarray}
\Delta_{qp}(J_0) & = & J_0 \> \otimes \> 1  +  1 \> \otimes \> J_0, \nonumber \\
\Delta_{qp}(J_3) & = & J_3 \> \otimes \> 1  +  1 \> \otimes \> J_3, \nonumber \\
\Delta_{qp}(J_{\pm}) & = & J_{\pm} \> \otimes 
\> {(qp)}^{{{1}\over{2}}J_0} \> {(qp^{-1})}^{+{{1}\over{2}}J_3} +
   {(qp)}^{{{1}\over{2}}J_0} \> {(qp^{-1})}^{-{{1}\over{2}}J_3} \>
\otimes \> J_{\pm}     
\label{eq:copro}  
\end{eqnarray}
% (11)
and is clearly seen to depend on the two parameters $q$ and 
$p$. [Note that with the constraint $p = q^{\ast}$, to be used in 
Subsec.~2.2, the co-product 
satisfies the Hermitean conjugation property               
 $(\Delta_{qp} (J_\pm))^\dagger = \Delta_{pq}(J_\mp)$
and is compatible with the commutation relations for the four operators
$\Delta_{qp} (J_\alpha)$ (with $\alpha = 0,3,+,-$).] 
The universal ${\cal{R}}$-matrix (for the coupling
of two angular momenta $I = \frac{1}{2}$) associated to the 
co-product $\Delta_{qp}$ reads 
\begin{equation} 
{\cal{R}}_{pq} = \pmatrix{
p&0&0&0\cr
0&\sqrt{pq}&0&0\cr
0&p-q&\sqrt{pq}&0\cr
0&0&0&p\cr
}, 
\label{eq:rmatrix}
\end{equation}
% (12)
and it can be proved that ${\cal{R}}_{pq}$ verifies the so-called 
Yang-Baxter equation.

The operator defined by$^{14}$ 
  \begin{equation}
 C_2( U_{ {qp}}({\rm u}_2))\  :=  \  {{1}\over{2}}(J_+J_- + J_-J_+)\  +\ 
 {{1}\over{2}} \> \qpn{2} \> (qp)^{J_0-J_3} \> 
\left( \qpn{J_3} \right)^2 
\label{eq:casim1} 
\end{equation}
% (13)
 is an invariant of the quantum algebra $U_{ {qp}}({\rm u}_2)$. It 
 depends truly on the two parameters $q$ and $p$. 
The invariant $C_2(U_{{qp}}({\rm u}_2))$ will be one of the main mathematical 
ingredients for the $qp$-rotor model to be developed below. 
Hence, it is worth to examine its structure more precisely, especially 
its dependence on two independent parameters. Equation (\ref{eq:copro}) 
suggests the following change of parameters 
\begin{equation}
Q := (qp^{-1})^{1\over 2} , \qquad   
P := (qp)     ^{1\over 2} .
\label{eq:paramet}
\end{equation}
% (14)
Then, by introducing the generators $A_\alpha$ ($\alpha = 0,3,+,-$) 
\begin{equation}
  A_0   \, := \, J_0 ,  \qquad \quad
  A_3   \, := \, J_3 ,  \qquad \quad
  A_\pm \, := \, (qp)^{-{1\over 2}(J_0 - {1\over 2})} \, J_\pm ,
\label{eq:Aoper}
\end{equation}
% (15) 
it can be shown that the two-parameter quantum algebra 
 $U_{{qp}}({\rm u}_2)$ is isomorphic to the central 
extension 
\begin{equation}
U_{qp}({\rm u}_2) = {\rm u}_1 \otimes U_Q({\rm su}_2), 
\label{eq:direct}
\end{equation}
% (16)
where $ {\rm u}_1$ is spanned by the operator $A_0$ and $U_Q({\rm su}_2)$ 
by the set $\{A_3, A_+, A_-\}$. The $Q$-deformation   $U_Q({\rm su}_2)$ 
(a one-parameter deformation!) 
of the Lie algebra  ${\rm su}_2$ corresponds to the usual commutation 
relations 
 \begin{equation}
[A_3,A_\pm   ] \ = \ \pm A_\pm ,  \qquad
[A_+,A_-     ] \ = \ [2A_3]_Q  .  
\label{eq:acomm}
\end{equation}
% (17)
Furthermore, the co-product relations (\ref{eq:copro}) leads to 
 \begin{equation}
\Delta_{qp}(J_{\pm}) \> = \> P^{ \Delta_{Q}(A_0) - \frac{1}{2} } 
\> \Delta_{Q}(A_{\pm}) ,  
\label{eq:cpfac}
\end{equation}
% (18)
where the co-product $\Delta_Q$ is given via 
\begin{eqnarray}
\Delta_{Q}(A_0) & = & A_0 \> \otimes \> 1 + 1 \> \otimes \> A_0 , \nonumber \\
\Delta_{Q}(A_3) & = & A_3 \> \otimes \> 1 + 1 \> \otimes \> A_3 , \nonumber \\
\Delta_{Q}(A_{\pm}) & = & A_{\pm} \> \otimes \> Q^{+A_3} +
                                                Q^{-A_3} 
                                  \> \otimes \> A_{\pm}.     
\label{eq:acopro}  
\end{eqnarray}
% (19)
Equations (\ref{eq:acomm}) involve only one parameter, i.e., the 
parameter $Q$. However, two parameters ($Q$ and $P$) occur 
in (\ref{eq:cpfac}) as well as in the invariant  $ C_2( U_{ {qp}}({\rm u}_2))$ 
transcribed in terms of $Q$ and $P$. As a matter of fact, 
(\ref{eq:casim1}) can be rewritten as
\begin{equation}
  C_2(U_{qp}({\rm u}_2)) \; = \;
  P^{2A_0-1} \; C_2(U_Q({\rm su}_2)) ,
\label{eq:inv1}
\end{equation}
% (20)
where
\begin{equation}
C_2(U_Q({\rm su}_2)) \; := \;
  {1 \over 2} \; (A_+A_- + A_-A_+) +
  {1 \over 2} \; [2]_Q \; \left( [A_3]_Q \right)^2
\label{eq:inv2}
\end{equation}
% (21) 
is an invariant of $U_Q({\rm su}_2)$ [compare Eqs.~(\ref{eq:casim1}) and 
                                                   (\ref{eq:inv2})]. 
As a consequence, of central importance for the $qp$-rotor model 
of Subsec.~2.2, the invariant $C_2(U_{qp}({\rm u}_2))$, in 
either the form (\ref{eq:casim1}) or the form (\ref{eq:inv1}), depends on 
two parameters. Finally, it 
should be noted that $C_2(U_{qp}({\rm u}_2))$ can be identified 
to the invariant of $U_q({\rm su}_2)$ and to the Casimir of $ {\rm su}_2 $ 
when $p = q^{-1}$ and $p = q^{-1} \to 1$, respectively.
In this sense, the $ U_{ {qp}}({\rm u}_2) $ symmetry 
encompasses the 
 $ U_{ {q}}({\rm su}_2)$  and  ${\rm su}_2 $ symmetries.

To close this section, let us say a few words on the 
representation theory of $U_{{qp}}({\rm u}_2)$ in the case 
where neither $q$ nor $p$ are roots of unity. An irreducible 
representation of this quantum algebra is described by a Young 
pattern 
$[\varphi_1; \varphi_2]$ with $\varphi_1 - \varphi_2 = 2I$, where $2I$
is a nonnegative integer ($I$ will represent a spin angular momentum in what
follows). We note
$\vet{[\varphi_1 ; \varphi_2]}{ M }$
(with $M = - I, -I+1,\cdots,+I)$
the basis vectors for the representation
$[\varphi_1 ; \varphi_2]$. 

We are now ready to develop a $qp$-rotor model for 
describing energy levels and reduced transition 
probabilities for rotation spectra of molecules 
and nuclei.   
 
 \subsection{Energy levels}
We want to construct a nonrigid rotor model. 
As a first basic hypothesis (Hypothesis 1), 
we take a rigid rotor with $U_{{qp}}({\rm u}_2)$ 
symmetry, thus introducing the nonrigidity through
the $qp$-deformation of the Lie algebra $u_2$. 
More precisely, 
we assume that the $qp$-rotor Hamiltonian $H$ 
is a 
linear function of the invariant $C_2( U_{ {qp}}({\rm u}_2))$: 
\begin {equation} 
H \; = \; { 1 \over 2{\cal I} } \; C_2 (U_{qp}({\rm u}_2)) + E_0,  
\label{eq:ham1} 
\end{equation}
% (22)
where ${\cal I}$  denotes the moment of inertia of the rotor and 
$E_0$ the bandhead energy. As a second hypothesis (Hypothesis 
2), we 
take $\varphi_1 = 2 I$ and 
     $\varphi_2 = 0$. This means that we work with 
state vectors of the type $\vet{I}{M} \equiv \vet{[2I;0]}{M}$. 
Therefore,   
the eigenvalues of $H$ are obtained by the action of $H$ on the 
physical subspace
$\{ \vet{I}{M} \ : \ M = -I, -I +1, \cdots, +I \}$
of constant angular momentum $I$ corresponding
to the irreducible
representation $[2I;0]$ of $U_{{qp}}({\rm u}_2)$. The two 
preceding hypotheses lead to the energy formula  
\begin{equation} 
  E(I)_{qp}  \; = \; { 1 \over {2 {\cal I}} } \;
  \qpn{I} \; \qpn{I+1} + E_0
\label{eq:eig1} 
\end{equation}
% (23)
for the $qp$-deformed rotational level of angular momentum $I$. 

By introducing $s = \ln q$ and $r = \ln p$, Eq.~(\ref{eq:eig1}) yields 
\begin{equation}
  E(I)_{qp} \; = \; { 1 \over {2{\cal I}} } \;
      {\rm e}^{ (2I-1) {{s+r} \over 2} } \;
      { {\sinh (I {{s-r}\over 2}) \;
         \sinh  [(I+1) {{s-r}\over 2}]} \over
       {\sinh^2 (  {{s-r}\over 2})} }
      + E_0 .
\label{eq:ener-rs1}
\end{equation}
% (24)
Preliminary studies have lead us to the conclusion that a good 
agreement between theory and experiment cannot be always obtained by 
varying  the parameters  
$s$ and $r$ (or $q$ and $p$) on the real line ${\bf R}$, 
a fact that confirms a similar 
conclusion reached in Ref.~20 for $p = q^{-1} \in {\bf R}$. In 
addition, if we want that our $qp$-rotor model reduces to the 
$q$-rotor model developed by Raychev, Roussev and Smirnov$^{19}$ 
when $p=q^{-1}$ (or equivalently $r = -s$), we are naturally 
left to impose that $(s+r)$ and $(s-r)/{\rm i}$ should be real numbers. 
[Observe that the two constraints $(s+r)         \in {\bf R}$
                              and $(s+r)/{\rm i} \in {\bf R}$ ensure 
that the energy $E(I)_{qp}$ is real as it should be.] 
Furthermore, 
work in progress in collaboration with J.~Meyer (in Lyon) shows
that for certain SD bands, a good 
agreement between theory and experiment requires that the 
parameters $s$ and $r$ vary on the real line ${\bf R}$. 
Thus, we shall consider the two possible parametrizations: 
\begin{eqnarray}
{\rm (a)} \qquad \qquad 
{{s+r}\over 2}         &   =   &  \beta \cos \gamma \in {\bf R}, \qquad
{{s-r}\over 2 {\rm i}}     =      \beta \sin \gamma \in {\bf R}, \nonumber \\
                       &       &                                           \\
{\rm (b)} \qquad \qquad                         
{{s+r}\over 2}         &   =   &  \beta \cos \gamma \in {\bf R}, \qquad
{{s-r}\over 2 {\rm i}}     =\frac{\beta \sin \gamma}{\rm i} 
                                            \in {\rm i} {\bf R}, \nonumber 
\label{eq:rs}
\end{eqnarray} 
% (25) 
so that the parameters $q$ and $p$ read 
 \begin{eqnarray}
{\rm (a)} \qquad \qquad
 q & = &           {\rm e}^{\beta \cos \gamma} \; 
          {\rm e}^{+{\rm i} \beta \sin \gamma} , \qquad  
 p   =   q^{\ast}= 
                   {\rm e}^{\beta \cos \gamma} \; 
          {\rm e}^{-{\rm i} \beta \sin \gamma}, \nonumber \\
   &   &                                                  \\ 
{\rm (b)} \qquad \qquad
 q & = &           {\rm e}^{\beta \cos \gamma} \; 
          {\rm e}^{+        \beta \sin \gamma} , \qquad  
 p   =             {\rm e}^{\beta \cos \gamma} \; 
          {\rm e}^{-        \beta \sin \gamma}. \nonumber
\label {eq:qpn1}   
\end{eqnarray}
% (26)
Thus, our $qp$-rotor model involves two independent real 
parameters $\beta$ and $\gamma$ 
corresponding  either to  (a) the two complex 
parameters $q$ and $p$ subjected to the constraint 
$p = q^{\ast}$     or to  (b) the two real 
parameters $q$ and $p$.  
In terms of the parameters $\beta$ and $\gamma$, 
the  energy formula (\ref{eq:ener-rs1}) takes the 
form 
\begin {eqnarray}
E(I)_{qq^{\ast}} &  = & { 1 \over {2 {\cal I}} } \;
      {\rm e}^{(2I-1) \beta \cos \gamma} \;
      { {\sin (I \beta \sin \gamma) \; \sin [(I+1) \beta \sin \gamma]} \over
        {\sin^2 (\beta \sin \gamma)} } 
      + E_0   \; \; \; \; \qquad  \qquad  
                      \quad   \quad 
(27 {\rm a}) \nonumber \\
{\rm or}      \qquad  \qquad  \quad & &          \nonumber \\
E(I)_{qp} \     &  = &  { 1 \over {2 {\cal I}} } \;
      {\rm e}^{(2I-1) \beta \cos \gamma} \;
      { {\sinh (I \beta \sin \gamma) \; \sinh [(I+1) \beta \sin \gamma]} \over
        {\sinh^2 (\beta \sin \gamma)} }
      + E_0   \qquad  \qquad  \quad \quad (27 {\rm b}) \nonumber 
\label{eq:eig2}
\end{eqnarray}
% (27)
in the parametrizations of type  (a) or (b), respectively. 

\addtocounter{equation}{+1} 

In the (a)-parametrization, 
to better understand the connection between our $qp$-rotor 
model and the $q$-rotor model of Ref.~19, we can perform a 
series analysis of Eq.~(27a). A straightforward 
calculation allows us to rewrite Eq.~(27a) as 
\begin{equation}
E(I)_{qq^{\ast}} \; = \; { 1 \over {2{\cal I}_{\beta \gamma}} } \;
      \left(
\sum_{n=0}^\infty \;
      d_n (\beta,\gamma) \;[I(I+1)]^n +
      (2I + 1) \;
\sum_{n=0}^\infty \;
      c_n (\beta,\gamma) \;[I(I+1)]^n \right) +
      E_0 ,
\label{eq:series}
\end{equation}
% (28) 
 where the expansion coefficients $d_n(\beta, \gamma)$
 and $c_n(\beta, \gamma)$
are given in turn by the series
\begin{eqnarray}
\lefteqn{ d_n(\beta, \gamma)  = 
 \frac{2^{2n-1}}{\sin^2(\beta \sin \gamma)} }   \nonumber  \\
  & &  \times  \somk \bigg\{ {(\cos \gamma)}^{2k+2n} \cos(\beta \sin \gamma)
- \cos[(2k+2n  ) \gamma] \bigg\}
 \frac{{\beta}^{2k+2n  }}{(2k+2n  )!} 
\frac{(k+n)!}{k! \ n!} ,  \nonumber \\
\lefteqn{c_n(\beta, \gamma)  =   \frac{2^{2n-1}}{\sin^2(\beta \sin \gamma)} }  \nonumber \\
   &  &  \times \somk \  \bigg\{ {(\cos \gamma)}^{2k+2n+1} \cos(\beta \sin \gamma)
- \cos[(2k+2n+1) \gamma] \bigg\} 
 \frac{{\beta}^{2k+2n+1}}{(2k+2n+1)!} 
\frac{(k+n)!}{k! \ n!} .     
\label{eq:termgene2} 
\end{eqnarray}
% (29) 
In Eq.~(\ref{eq:series}), we have introduced the deformed moment of inertia:
\begin{equation}
{\cal I}_{\beta \gamma} = {\cal I} \, {\rm e}^{2 \beta \cos \gamma} ,
\label{eq:inertie1}
\end{equation}
% (30)
which gives back the ordinary moment of inertia when 
$\gamma = \pi/2$ (i.e., 
$q = p^{-1} = {\rm e}^{{\rm i} \beta}$). In the limiting 
situation where 
$\gamma = \pi/2$, the coefficients 
$c_n(\beta, \gamma)$
vanish and the energy formula (\ref{eq:series}) simplifies to 
\begin{equation}
E(I)_{qq^{-1}} \; = \; {1 \over 2 {\cal I}} \;
      {{\beta^2} \over {\sin^2 \beta}} \;
      \sum_{n=1}^\infty \; (-1)^{n-1} \;
      {{2^{n-1}} \over {n!}} \;
      \beta^{n-1} \; j_{n-1}(\beta) \; [I(I+1)]^n + E_0,
\label{eq:bonat1}
\end{equation}
% (31)
where $j_{n-1}$ denotes a spherical
Bessel function of the first kind. Equation (\ref{eq:bonat1}) 
was derived by 
Bonatsos {\em et al}.$^{20}$ for the $q$-rotor model with 
$q= {\rm e}^{{\rm i} \beta}$ in order to prove the mathematical 
parentage between the $q$-rotor model and the variable moment 
of inertia (VMI) model.$^{40}$ The 
series (\ref{eq:bonat1}) corresponds indeed to the compact expression 
\begin{equation}
E(I)_{qq^{-1}} \; = \; {1 \over 2 {\cal I}} \;
      [I]_q [I+1]_q  + E_0 ,
\label{eq:bonat2}
\end{equation}
% (32)
to be compared with Eq.~(\ref{eq:eig1}). 
Note that Eq.~(32) corresponds also to the 
(b)-parametrization with $\gamma = \pi / 2$. 
The transition from 
                    Eq.~(\ref{eq:eig1}) to 
                    Eq.~(\ref{eq:bonat2}) 
illustrates the descent from the $U_{qp}({\rm u}_2)$ symmetry 
of the $qp$-rotor to the $U_{q}({\rm su}_2)$ symmetry of the 
$q$-rotor. A further descent in 
symmetry is obtained when $\beta \to 0$ 
              (i.e., $q = p^{-1} \to 1$): in this case 
$[I]_q [I+1]_q \ \to \  I (I + 1) $ and we get 
[from Eq.~(\ref{eq:bonat2})] 
the usual energy 
formula corresponding to the rigid rotor with ${\rm su}_2$ 
symmetry. 
 
\subsection{E2 reduced transition probabilities}
 We now examine the implication of the 
 $U_{qp}({\rm u}_2)$ symmetry on the calculation of the 
 (electric quadrupole) 
 reduced 
 transition probability $B({\rm E2}; I+2 \to I)$. 
 Let us start with the ordinary expression of the reduced 
transition probability, namely, 
 \begin{equation}
B({\rm E2}; I+2  \to I)  =  \frac {5} {16 \pi}\; Q_0^2\; 
 { \bigg  \vert { \cg{I+2}{M}{2}{0}{I+2}{2}{I}{M}} \bigg  \vert} ^2 
\label{eq:be2n1}
\end{equation}
% (33)
for an E2 transition.$^{41}$ 
In Eq.~(\ref{eq:be2n1}), $Q_0$ is 
 the intrinsic electric quadrupole moment in the body-fixed frame. 
The coefficient of type $\cg{j}{m}{k}{\mu}{j}{k}{j'}{m'}$ in 
the right-hand side of Eq.~(\ref{eq:be2n1}) is a usual 
Clebsch-Gordan coefficient for the group SU(2). Our goal is to 
find a $qp$-analog of $B({\rm E2}; I+2 \to I)$.  
The strategy for obtaining a $qp$-analog of Eq.~(33) is the following: 

{(i)} We first rewrite the SU(2) Clebsch-Gordan coefficient of 
Eq.~(\ref{eq:be2n1}) in terms of a matrix element of an SU(2) 
unit tensor operator $t_{k\mu\alpha}$ with $k=2$, $\mu=0$ and 
$\alpha=-2$. This may be done from the    general formula$^{38}$ 
 \begin{equation}
  \vra{j'}{m'} t_{k \mu \alpha} \vet{j}{m} = \delta(j',j+\alpha) 
                                             \delta(m',m+\mu   ) (-1)^{2k} 
(2j'+1)^{- \frac{1}{2}} \;  \cg{j}{m}{k}{\mu}{j}{k}{j'}{m'} ,  
\label{eq:t202}
\end{equation} 
% (34)
 which shows that the  irreducible  tensor 
 operator $t_{k\mu\alpha}$  produces
the  (angular momentum) state vector ${\vet {j+ \alpha}{m+  \mu}}$  
  when acting upon the  state vector ${\vet {j}        {m      }}$. 
Then, Eq.~(\ref{eq:be2n1}) is amenable to the form 
 \begin{equation}
B({\rm E2}; I+2  \to I)    =   \frac {5}{16 \pi}\; {Q}_0^2 \; 
 (2I+1 )\;   {\bigg  \vert
\vra{I}{M} t_{20-2} \vet{I+2}{M} \bigg  \vert }^2 
\label{eq:be2no1}
\end{equation} 
% (35)
by making use of Eq.~(\ref{eq:t202}). 

{(ii)} We know that the general operator  $t_{k\mu\alpha}$   
 can be  realized  
 in terms of  two pairs  $\{ {a}_+^+ ,{a}_+ \}$ and
                         $\{ {a}_-^+ ,{a}_- \}$     
of ordinary boson operators.    In this respect, we 
may consider the so-called van der Waerden$^{38}$ realization of 
$t_{k\mu\alpha}$. 
There are several ways to $qp$-deform the operator $t_{k\mu\alpha}$. 
Here, we choose to define a $qp$-deformation $t_{k \mu \alpha}(qp)$ 
by replacing, in the van der Waerden realization of $t_{k \mu \alpha}$,
the ordinary bosons 
$\{ {a}_{\pm}^+ ,{a}_{\pm} \}$ by $qp$-deformed bosons 
$\{ \ti{a}_{\pm}^+ ,\ti{a}_{\pm} \}$ and the ordinary 
factorials $x!$ by $qp$-deformed factorials 
$\qpn{x}! \> = \> \qpn{x  } \> 
                  \qpn{x-1} \> 
                  \cdots    \> 
                  \qpn{1  }$ 
for $x \in {\bf N}$. We thus obtain
\begin{eqnarray}
 t_{k \mu \alpha}(qp) & = & (-1)^{k+\alpha} \bigg ( {\frac{\qpn{k+\mu}! \; \qpn{k-\mu}! \; \qpn{k+\alpha}! \;
 \qpn{k-\alpha}! \;  \qpn{2j - k+\alpha}!}{ \qpn{2j + k+\alpha+1}!}}\bigg )^{\frac{1}{2}} \nonumber \\
                                       &      & \times  \sum_z (-1)^z {\frac{(\ti{a}_-^+)^{k- \mu -z} 
  (\ti{a}_-)^{k - \alpha  -z} (\ti{a}_+^+)^{ \mu + \alpha +z} (\ti{a}_+)^{z}}
{ \qpn{k-\mu -z}! \;  \qpn{k- \alpha - z}! \;  \qpn{\mu + \alpha + z}! \;  
\qpn{z}! } } .  
\label{eq:opu2} 
\end{eqnarray}
% (36) 
In particular, the $qp$-deformed operator $t_{20-2}(qp)$ connecting 
the state vector $\vert I+2,M)$, with $j  \equiv I+2$, to 
the state vector $\vert I  ,M)$, with $j' \equiv I  $, reads  
 \begin{equation}
 t_{20-2}(qp)  =  \bigg ( \frac{\qpn{3} \;  
                                \qpn{4} \; 
                                \qpn{2I}!} 
                               {\qpn{2} \; 
                                \qpn{2I +5}!} \bigg )^{\frac{1}{2}} 
    (\ti{a}_+)^{2}  (\ti{a}_-)^{2} , 
\label{eq:opu202} 
\end{equation} 
% (37)
an expression 
of direct interest for deriving the $qp$-analog of $B({\rm E2}; I+2  \to I)$. 

{(iii)} We assume that the $qp$-analog 
$B({\rm E2}; I+2  \to I)_{qp}$ of $B({\rm E2}; I+2  \to I)$ is simply 
 \begin{equation}
B({\rm E2}; I+2  \to I)_{qp}   :=   \frac {5} {16 \pi}\; Q_0^2 \;  
   \qpn{2I+1}\;   {\bigg  \vert
\vra{I}{M} t_{20-2}(qp) \vet{I+2}{M} \bigg  \vert }^2 .
\label{eq:qpbe2n2}
\end{equation}
% (38)
[Equation (\ref{eq:qpbe2n2}) constitutes the third and last 
hypothesis (Hypothesis 3) for our $qp$-rotor model.] By using 
Eqs.~(\ref{eq:opu202}) and (\ref{eq:qpbo2}), 
the relevant matrix element of the 
operator $t_{20-2}(qp)$ is easily found to be 
  \begin{eqnarray}
\lefteqn{\vra{I}{M} t_{20-2}(qp) \vet{I+2}{M}  = }  \nonumber \\
&  &  \bigg ( \frac{            \qpn{3}     \;  
                                \qpn{4}     \;
                                \qpn{2I}!   \; 
                                \qpn{I+M+1} \; 
                                \qpn{I-M+1} \; 
                                \qpn{I+M+2} \; 
                                \qpn{I-M+2} }
                                {\qpn{2}    \; 
                                \qpn{2I +5}!}  \bigg )^{\frac{1}{2}} .  
                                                    \nonumber \\
\label{eq:matx}
\end{eqnarray}
% (39)
 Then, the introduction of Eq.~(\ref{eq:matx}) into Eq.~(\ref{eq:qpbe2n2})
yields  
\begin{equation}
 B({\rm E2}; I+2  \to I)_{qp}    =  { \frac {5} {16 \pi}}\;  Q_0^2 \; 
 \frac{\qpn{3} \; \qpn{4}}{\qpn{2}}
\frac{\big(\qpn{I+1} \; \qpn{I+2}\big)^2}{\qpn{2I +2} \; \qpn{2I+3} \; 
\qpn{2I+4} \; \qpn{2I+5}} 
\label{eq:qpbe2n3}
\end{equation}
% (40)
in the case of the $K \equiv M = 0$ bands. 

At this stage, a contact with the formula 
 $ B({\rm E2}; I+2  \to I)_{q} $ derived by 
 Raychev, Roussev and Smirnov$^{19}$ 
 is in order. First, by taking $p=q^{-1}$ the 
right-hand side of (\ref{eq:qpbe2n3}) may be specialized to the 
expression of $ B({\rm E2}; I+2  \to I)_{q} $ obtained in Ref.~19. 
Hence, our $qp$-rotor model for the E2 reduced transition probability 
admits as a particular case the corresponding $q$-rotor 
model worked out in Ref.~19. Second, it can be 
shown that  
 \begin{equation}
  B({\rm E2}; I+2  \to I)_{qp}  =  P^{-4(I+1)}
  B({\rm E2}; I+2  \to I)_{Q },  
\label{eq:contact}
\end{equation}
% (42)
where $P$ and $Q$ are given by (\ref{eq:paramet}).  

Let us close with a remark. 
Should we have chosen to find a $qp$-analog of the 
Clebsch-Gordan coefficient in (\ref{eq:be2n1}), we would have 
obtained$^{39}$ 
 \begin{equation}
{\cg{I+2}{M}{2}{0}{I+2}{2}{I}{M}}_{qp}
=
{\cg{I+2}{M}{2}{0}{I+2}{2}{I}{M}}_{Q }
\label{eq:KAS}
\end{equation}
% (43)
and, consequently 
 \begin{equation}
  B({\rm E2}; I+2  \to I)_{qp}  =  
  B({\rm E2}; I+2  \to I)_{Q }. 
\label{eq:KASCON}
\end{equation}
% (44)
We prefer to use (\ref{eq:contact}) rather than 
                 (\ref{eq:KASCON})
because the factorization in (\ref{eq:contact}) parallels the 
one in (\ref{eq:inv1}). 

\section{Conclusions}
In this communication, we concentrated 
on a $qp$-rotor system 
[in the framework 
of a revisiting of the two-parameter quantum 
algebra $U_{ qp}({\rm u}_2)$]
that leads to a nonrigid rotor model (the $qp$-rotor model) 
based on three hypotheses.
The two facets of this model 
consist of a 
 three-parameter energy level formula 
and 
a $qp$-deformed E2 reduced transition probability formula. 
As limiting cases, the $qp$-rotor model gives back
the $q$-rotor model$^{19}$ (when $p = q^{-1}$) 
based on the quantum algebra $U_{ q}({\rm su}_2)$ and 
the rigid rotor model (when $p = q^{-1} \to 1$)
based on the Lie algebra su$_2$. 

A work to be published elsewhere$^{42}$ shows 
that the $qp$-rotor model is appropriate for  
describing the collective phenomenon of distortion 
occurring in the rotation of the nucleus 
(increase or decrease of the dynamical moment of inertia with the spin). 
The net difference 
between the $q$- and $qp$-rotor models comes from the ``quantum
algebra''-type parameter $a$ that tends to smooth the 
(spherical or hyperbolical) sine term in the energy
and thus accentuates or moderates the distortion phenomenon of the nucleus. 

To close this work, let us mention that Hypothesis 2 (i.e., 
$\varphi_1 = 2I$ and $\varphi_2=0$) 
of our model might be abandoned. This would lead to a {\em \`a la} 
Dunham formulation
for describing 
more complicated rotational spectra of 
deformed and superdeformed nuclei or 
rovibrational spectra of diatomic molecules.
As a further extension, it would be also interesting to combine our model with
one of Ref.~24 (based on the $q$-Poincar\'e symmetry) in the case of heavy 
nuclei. Work in these directions is in progress.     
 
\vskip 1 cm

\noindent {\bf Acknowledgments}

\noindent The authors would like to thank J.~Meyer 
and Yu.~F.~Smirnov for valuable discussions. They are also 
indebted to M.~Tarlini for calling their attention to 
Ref.~24. This work was presented by one of the authors (M.~K.) 
to the International Workshop ``Finite Dimensional Integrable 
Systems'' (Dubna, Russia, July 1994). 

\vskip 1 true cm 
 
\baselineskip 0.50 true cm

\end{document}